# Observation of Photoluminescence from a Natural van der Waals Heterostructure

Viviane Z. Costa,[1] Bryce Baker,[1] Hon-Loen Sinn,[1] Addison Miller,[1] K. Watanabe,[2] T. Taniguchi,[2] Akm Newaz[1]

Date: February 25th, 2022

[1]Department of Physics and Astronomy, San Francisco State University, San Francisco, California 94132, USA

[2] National Institute for Materials Science, Namiki 1-1, Tsukuba, Ibaraki 305-0044, Japan

Van der Waals heterostructures comprised of two-dimensional (2D) materials offer a platform to obtain materials by design with unique electronic properties. Franckeite (Fr) is a naturally occurring van der Waals heterostructure comprised of two distinct alternately stacked semiconducting layers; (i) $SnS_2$ layer and (ii) $Pb_3SbS_4$. Though both layers in the heterostructure are semiconductors, the photoluminescence from Franckeite remains elusive. Here, we report the observation of photoluminescence (PL) from Franckeite for the first time. We observed two PL peaks at ~ 1.93 eV and ~ 2.11 eV. By varying the temperature from 1.5 K to 80 K, we found that the PL peak position redshifts and the integrated intensity decreases slowly as we increase the temperature. We observed linear dependence of photoluminescence integrated intensity on excitation laser power indicating that the photoluminescence is originating from free excitons in the $SnS_2$ layer of Fr. By comparing the PL from Fr with the PL from a monolayer $MoS_2$, we determined that the PL quantum efficiency from Fr is an order of magnitude lower than that of a monolayer $MoS_2$. Our study provides a fundamental understanding of the optical behavior in a complex naturally occurring van der Waals heterostructure, and may pave an avenue toward developing nanoscale optical and optoelectronic devices with tailored properties.

Van der Waals (vdW) heterostructures prepared by stacking dissimilar two-dimensional (2D) materials such as transition metal dichalcogenides (TMDs), Graphene (Gr), or hexagonal boron nitride (hBN) has become a major research direction in material science and condensed matter physics.[1,2] These vdW heterostructures demonstrate novel electronic, optical, and optoelectrical properties that may find applications from nanoelectronics to nano-optics. These lab-prepared vdW heterostructures demonstrate electronic, optical, and thermal properties that strongly differ from those of the constituent 2D materials, thus opening the door to obtain on-demand interesting physical, electrical, optical, optoelectrical behavior, and thermal properties, such as moiré exciton,[3] and strongly-correlated quantum phenomena, including tunable Mott insulators at half-filling,[4] unconventional superconductivity,[5] and ferromagnetism.[6] The standard fabrication method of all these heterostructures includes manual or robotic vertical assembly of 2D stacks using deterministic placement methods.[2,7] Because of manual stacking in a laboratory setting, the neighboring layer interface may contain fabrication artifacts, such as foreign particles or bubbles between the interfaces or crystal mismatch, which may affect the measurement of intrinsic physical properties.[8-11] On the other hand, Fr provides a platform to measure intrinsic physical properties of interfaces with van der Waals heterostructures free of fabrication artifacts. We present here the optical and optoelectrical behavior of Franckeite, which is a naturally grown van der Waals superlattices layered semiconductors and free from fabrication artifacts between semiconducting layers.[8-12]

Naturally occurring Franckeite is a van der Waals heterostructure composed of alternating sequences of weakly bound stacked PbS-like pseudotetragonal $Pb_3SbS_4$ (Q) layers and $SnS_2$ pseudohexagonal (H) layers attached by van der Waals interactions.[8-10] Recently, several groups have demonstrated exfoliation of Franckeite (mechanically and by liquid-phase exfoliation) down to the single unit cell and the exfoliated flakes have been assembled into electronic devices, photodetectors operating in the near-infrared range, and energy conversion devices.[8-10,12]

The photoluminescence from a semiconductor is a fundamental property and plays a key role in understanding the physical mechanism in a semiconductor and is important for any optical or optoelectrical device applications. Though Fr provides a fascinating 2D semiconductor heterostructure with possibilities of exploring new physics and developing many attractive applications, the photoluminescence of Fr remains elusive. Here we demonstrate the first observation of photoluminescence from Fr. We have observed a strong signature of PL at varying temperatures from 1.5 K to 80 K. To understand the origin of the PL, we have also fabricated electrically-connected Fr devices and measured the photocurrent spectroscopy. The observed energy of the PL peak indicates that the PL is originating from $SnS_2$ layer. Moreover, the linear excitation power dependence of PL suggests that the PL is caused by the free excitons in $SnS_2$. By comparing the PL from Fr with the PL from a monolayer $MoS_2$ (1L-$MoS_2$), we showed that the PL quantum efficiency in Fr is an order of magnitude lower than that of 1L-$MoS_2$.

**RESULTS AND DISCUSSIONS**

We present our experimental study of PL and photocurrent spectroscopy of thermodynamically stable Franckeite immobilized flakes on a TEM grid and glass substrate, respectively. The bulk Fr crystals were obtained from San Jose mine in Bolivia. We have prepared the sample by using micro-exfoliation from bulk samples. First, we microexfoliated Fr sample on $SiO_2$ (90 nm)/$p^+$Si substrate. The thickness of the Fr samples is ~ 60- 100 nm.

After identifying an appropriate flake, Polyethylene terephthalate (PET) stamps were employed to pick up the Fr from the $SiO_2$ substrate at 60 °C. The Fr flakes were then released onto a TEM grid at 90–110 °C, followed by dissolving PET residues in dichloromethane for at least 12 hours. Since the samples reported here and the samples reported in an earlier publication are microexfoliated from the same bulk crystal, we point the reader to the work by Ray *et al.*, for a detailed characterization of the Franckeite by energy-dispersive X-ray (EDX) and scanning electron microscope[10] and detail study of vibrational properties of Fr.[13] Also, detail characterizations and high-resolution transmission electron microscope images of Fr are reported in the pioneering work by Velický[9] *et al.*, and Molina-Mendoza *et al.*[8]

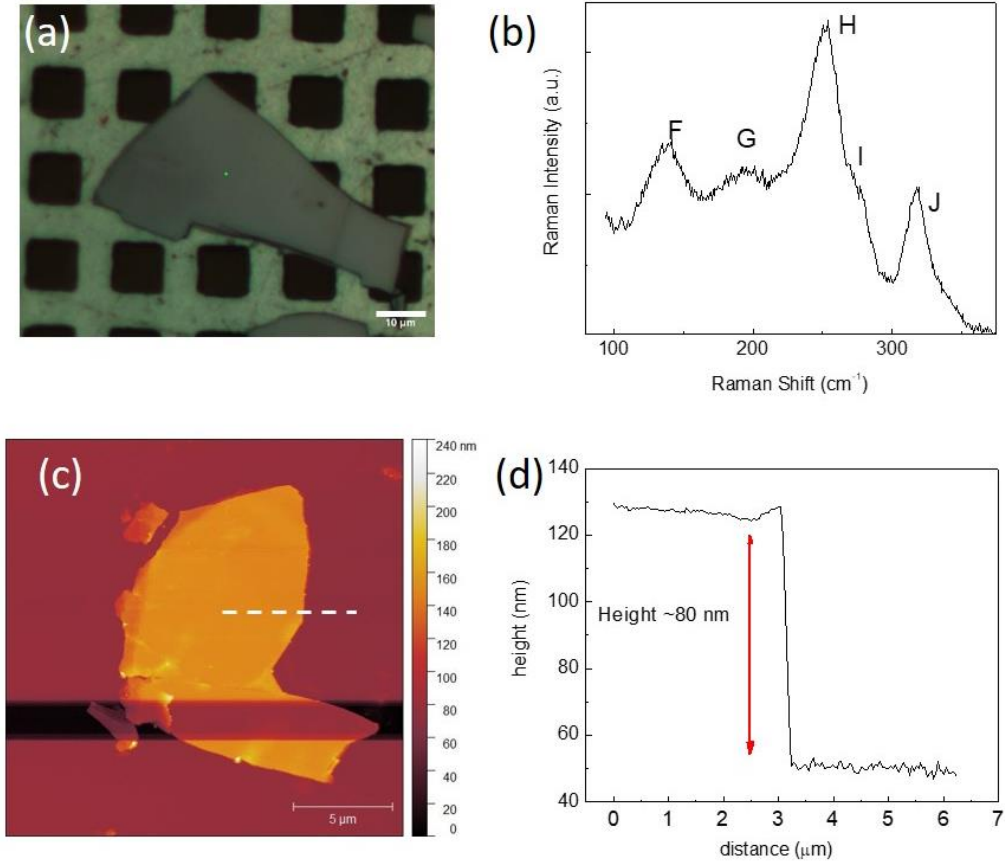

Figure 1: (a) The optical image of a Franckeite sample immobilized on a TEM mesh grid. The thickness of the sample is ~100 nm. The scale bar is 10 μm. (b) The AFM image of a Fr flake on SiO2/Si substrate before transferring to the TEM grid using dry transfer technique. (c) the height profile of the AFM image along the dashed white line in Fig.(b). The thickness of the Fr flake is ~ 80 nm. (d) Raman spectrum of the sample recorded at room temperature. The laser excitation wavelength was 633 nm with power ~70 μW.

We studied Fr sample on top of a TEM grid to avoid any photogating effect arising from the substrate.[14] Fig.1(a) presents the optical image of a Fr sample immobilized on top of a TEM grid. We also studied Fr sample on a $SiO_2$(90nm)/p$^+$-Si substrate and obtained a similar PL spectrum (see Supporting Information). In this manuscript, we are presenting the results only from the sample residing on TEM grids. Fig.1(b) presents the AFM image of a Fr flake. Fig.1(c) presents the height profile along a white dashed line, which shows that the thickness of the flake is 80 nm.

Fig.1(d) presents the room temperature Raman spectroscopy from the Fr flakes. Confocal micro-Raman measurements were performed using commercial equipment (Horiba LabRAM Evolution). A long working-distance 100× objective lens with a numerical aperture of 0.6 was used. The excitation source was a 633 nm laser of power ~200 μW. The Raman spectra are measured using a grating with 1200 g/mm blazed at 500 nm and a solid-state-cooled CCD detector. We have observed several Raman modes, as F(140 cm$^{-1}$), G(195 cm$^{-1}$), H(256 cm$^{-1}$), I(276 cm$^{-1}$), and J(321 cm$^{-1}$), which are the characteristic signatures of Fr.[13]

Fig.2(a) presents the PL spectrum from a Fr flake measured at different temperatures from 1.5 K to 80 K. We measured the PL spectroscopy with a continuous-wave excitation laser at 532 nm. Cryogenic confocal photoluminescence spectroscopy was performed in an ultralow-vibration closed-cycle cryostat (attocube systems, attoDRY2100) with a base temperature of 1.5 K. The Fr samples were positioned with x-y-z nano-positioners into the focal plane of a low-temperature apochromat with a numerical aperture of 0.85 (attocube systems, LT-APO/Raman/0.85). The PL signal was collected and guided by a multi-mode fiber into a spectrometer (Andor Shamrock 350i) equipped with a thermoelectrically cooled CCD camera. The laser power on the sample was ~ 1 mW measured by a power meter.

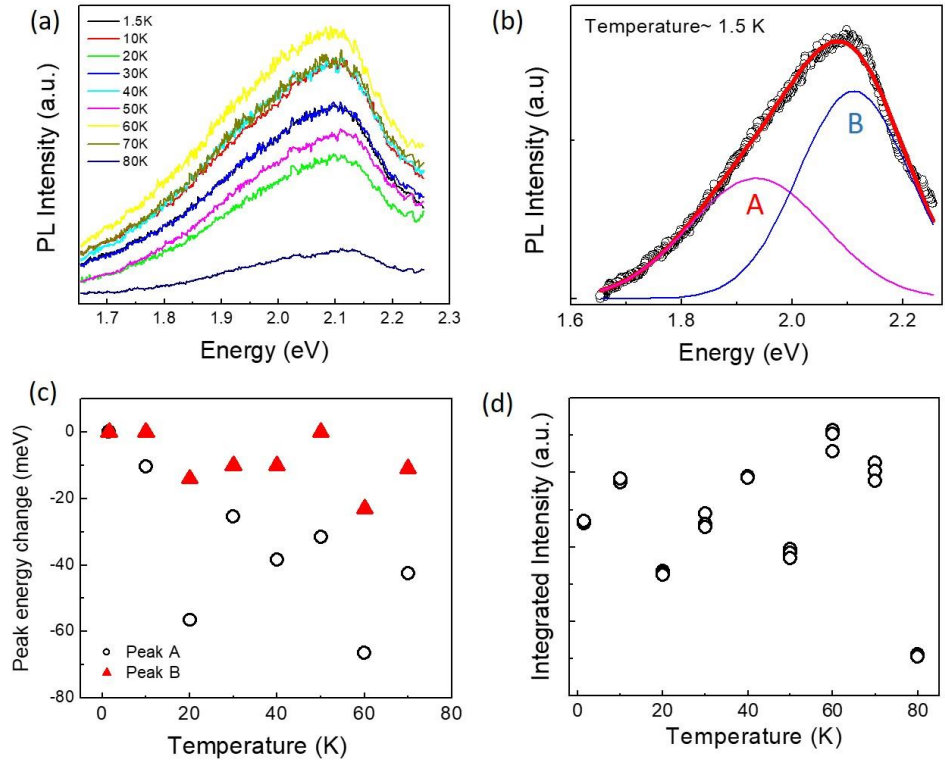

Figure 2: (a) Photoluminescence intensity from the Fr flake at varying temperature from 1.5 K to 80K. The laser excitation wavelength is 532 nm. (b) PL spectrum measured at ~1.5 K (black circles). The data were fitted but two gaussian peaks (magenta and black lines). The two peaks are marked by A- and B-peak. The resultant fit lines are shown by red line. (c) The plot presents the temperature-dependent peak positions of A- (black circle) and B-peaks (red triangles) measured by the Gaussian fitting shown in Fig.(b). (d) Integrated photoluminescence intensity from Fr at varying temperature. Multiple data points at a single temperature means that the data were obtained a multiple time at that temperature.

We observed a broad peak with a peak at 2.1 eV as shown in Fig.2(a). But the peak is very asymmetric with a long tail in the lower energy suggesting that the observed peak is a convolution of two peaks. Indeed, the peak can be fit very well, by two Gaussian peaks as shown in Fig.2(b). The peak is fit poorly with a single peak (see Supplementary Information) confirming that the PL spectrum is a convolution of two peaks. Moreover, we see the signature of two peaks for PL in Fr sample residing on SiO2/Si+p substrate (see Supplementary

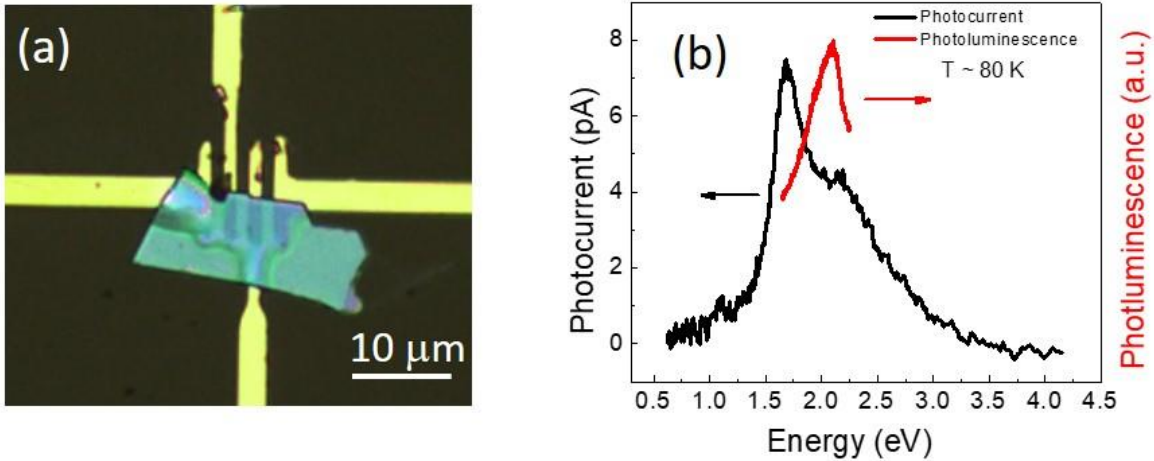

Figure 3: Comparison between photoluminescence spectroscopy and photocurrent spectroscopy. (a) The optical image of a thin Fr flake, which is electrically connected by Au electrodes. The yellow lines are the Au metal electrodes. See text for details. (b) Photocurrent spectroscopy data and Pl data from Franckeite measured at 80 K. The PL data was recorded using a cryogenic optical objective with a working distance ~100 um. The photocurrent data was measured using a different Fr flake of similar thickness.

Information). We have named the lower energy peak as A-peak and the higher energy peak as the B-peak. For the PL data measured at 1.5 K, we found that the A- and B-peaks are at 1.93 eV and 2.11 eV, respectively.

Fig.2(c) presents the peak position changes ($\Delta = E_{1.5} - E_T$, where $E_{1.5}$ and $E_T$ are the peak position at 1.5 K and a temperature $T$, respectively). We see both peaks redshift as we increase the temperature. But A-peak redshifts at a higher rate with respect to temperature than the B-peak. Fig.2(d) presents the integrated intensity values for that temperature range. We observed that the integrated intensity has a weak dependence on the temperature. The integrated intensity decreases slowly as we increase the temperature.

To understand the origin of the PL peak from Fr, we also prepared electrically connected Fr devices and measured photocurrent spectroscopy. First, we fabricated the metal electrodes on a glass substrate using optical lithography followed by thermal evaporation of Cr/Au (5 nm/100 nm). We selected a glass substrate for our devices to avoid the effect of the photovoltage or photogating on the optoelectronic behavior of our samples.[14] In parallel, we prepared a PDMS (poly (dimethylsiloxane)) stamp structure on a microscope slide. The Fr was exfoliated onto the PDMS stamp followed by characterization using optical microscopy, Raman spectroscopy, and atomic force microscopy to verify the Fr deposition and determine the sample thickness. We aligned the exfoliated Fr flake on the polymer stack with the pre-patterned metal electrode target under a microscope and brought it into contact. The polymer layer was mechanically separated from the PDMS stacks. The optical image of one electrically connected Fr device is shown in Fig.3(a), where the yellow lines are the Au electrodes.

The sample was mounted inside a microscopy cryostat (Janis Research, ST-500) equipped with electrical feedthrough for electro-optical measurements. We used a broadband light source (tungsten–halogen lamp) coupled to a double-grating monochromator (Acton Spectra Pro SP-2150i). The photocurrent was measured by employing lock-in techniques. Detail study of the optoelectrical properties of Fr is available in our previous publications.[10]

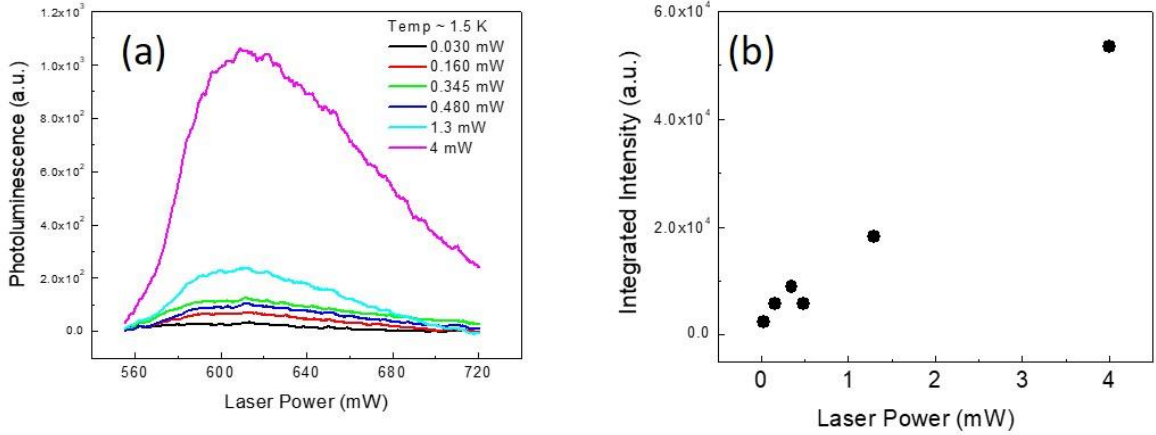

Figure 4: Photoluminescence intensity from Fr for different laser power from 30 µW to 4 mW. The laser excitation wavelength is 532 nm. The laser beam diameter was ~1 micron. (b) Integrated photoluminescence intensity from Fr for different laser power. The PL intensity varies linearly with respect to the laser power.

The photocurrent spectrum and PL spectrum of Fr measured at 80 are presented in Fig.3(b). We observed two peaks in the photocurrent spectroscopy. The energy of the two peaks in photocurrent spectroscopy, ~590 nm (~2.1 eV) and ~740 nm (~1.66 eV), match the optical band gaps of $SnS_2$ and $Sb_2S_3$, respectively.[15] Interestingly, we observed only a broad asymmetric peak in PL near ~2.1 eV. This suggests that the PL is mostly dominated by the excitons from $SnS_2$ layer.

To understand the origin of the PL in Fr, we conducted an excitation intensity-dependent PL study with power varying from 30 µW to 4 mW, i.e., more than two orders of power variations as shown in Fig.4. The behavior of the integrated PL intensity depending on the laser excitation power is considered a good indicator of the nature of the radiative recombination processes giving rise to the different spectral features near the band edge.[16] If the integrated PL intensity ($I$) power law depends on the laser excitation power, $L$, as $I \propto L^k$, a value of $k \sim 1$ implies a free exciton-like transition and $k < 1$ suggests a excitons involving defects/impurities such as free-to-bound (recombination of a free hole and a neutral donor or that of a free electron and a neutral acceptor) and donor-acceptor pair transitions.[16]

Fig.4(a) shows the power-dependent PL measurements for different laser excitation power measured at 1.5 K. The laser power was varied by neutral density filter and the power was measured by a Si power meter (Thorlabs-PM 100D). The integrated PL intensity for different laser power is shown in Fig.4(b). We observed a linear power dependence behavior, which suggests that the PL is originating from the recombination of free excitons in $SnS_2$.

To measure the PL efficiency qualitatively, we also measured the PL from a monolayer $MoS_2$ sample residing on hBN substrate in the same setup. The samples were fabricated by sequential transfer by using PET stamps. We have selected 1L-$MoS_2$ as it is a direct band-gap semiconductor at the monolayer limit, demonstrates high PL efficiency, and has become a benchmark of PL study in atomically thin 2D semiconductors.[17] The PL from both Fr and 1L-$MoS_2$ is shown in Fig.5. We observed the signatures of neutral excitons ($A^0$ and B) as well as charged excitons ($A^-$ excitons or excitons composed of two electrons and one hole).[17] To compare the PL efficiency properly we normalized the PL count by power and exposure time. We found that the PL efficiency of Fr needs to be multiplied by a factor of 10 to be comparable to the PL from a monolayer $MoS_2$. Hence, the PL from Fr is an order of magnitude lower than 1L-$MoS_2$ as shown in Fig.5.

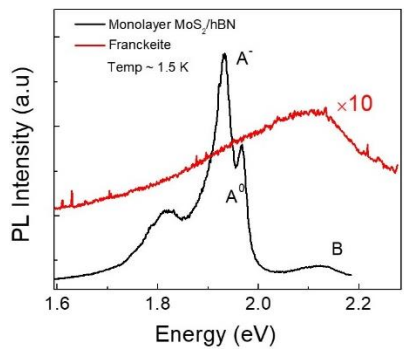

Figure 5: Comparison of Fr PL with a monolayer $MoS_2$ sample. The PL was measured at 1.5 K. The PL from Fr is multiplied by a factor of 10 to make the data comparable with MoS2 PL. The neutral exciton, charged excitons and B-excitons are marked by $A^0$, $A^-$, and B, respectively.

In conclusion, we observed the photoluminescence from Fr, a natural van der Waals heterostructure, with a PL peak around 2.1 eV. We found that the PL is originating from the free excitons in the pseudohexagonal $SnS_2$ layer of Fr. We also demonstrate the efficiency of PL from Fr is an order of magnitude lower than the efficiency of PL from direct-band monolayer $MoS_2$. Our study demonstrates the PL from a natural van der Waals heterostructure that may help understand the optical and optoelectrical properties of many complex heterostructures and develop next-generation optical and optoelectrical devices.

**Acknowledgement**

G.B., V.Z.C., N. B., and A.K.M.N. acknowledge the support from the Department of Defense Award (ID: 72495RTREP). A.K.M.N. also acknowledges the support from the National Science Foundation Grant ECCS-1708907 and the faculty start-up grant provided by the College of Science and Engineering at San Francisco State University. All AFM measurements were supported by NSF for instrumentation facilities (NSF MRI-CMMI 1626611). All low photoluminescence measurements were conducted using an attocube cryostat that was supported by NSF for instrumentation facilities (NSF DMR- 1828476). All Raman spectroscopy data were acquired at the Stanford Nano Shared Facilities (SNSF), supported by the National Science Foundation under award ECCS-2026822.

Supplementary Materials for

# Observation of Photoluminescence from a Natural van der Waals Heterostructure

Viviane Z. Costa,[1] Hon-Loen Sinn,[1] Bryce Baker,[1] Addison Miller,[1] K. Watanabe,[2] T. Taniguchi,[2] Akm Newaz[1]

[1]Department of Physics and Astronomy, San Francisco State University, San Francisco, California 94132, USA

[2] National Institute for Materials Science, Namiki 1-1, Tsukuba, Ibaraki 305-0044, Japan

1) Photoluminescence of Fr on a SiO2/Si substrate

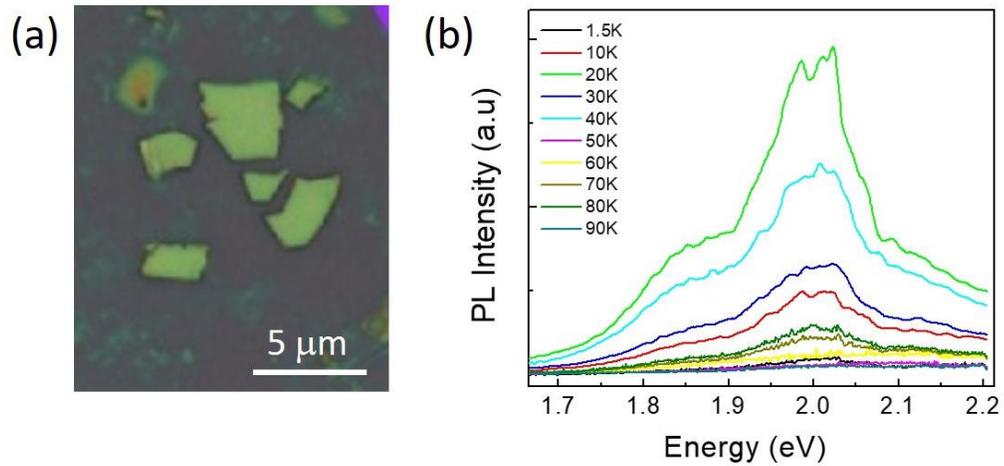

Figure 1: (a) The optical image of Franckeite flakes immobilized on SiO2/Si substrate. (b) The plots shows the PL measured at different temperature from 1.5 K to 90 K.

2) Single Gaussian fitting of the PL spectrum

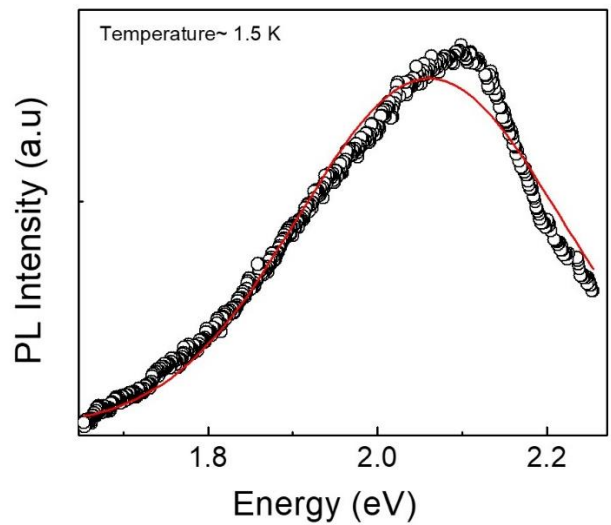

Figure 2: The black circle presents the measured Pl spectrum data at 1.5K. The red solid line shows the Gaussian fit.